\documentclass[
 preprint,
 amsmath,amssymb,
 aps,
 pre,
]{revtex4-2}

\usepackage{graphicx}
\usepackage{bm}
\usepackage{booktabs}

\newtheorem{theorem}{Theorem}
\newtheorem{lemma}[theorem]{Lemma}

\begin{document}

\title{Decision-Related Cognitive Signatures from Fast-Slow Dynamics: A Low-Dimensional Observation-Operator Framework}

\author{Furkan Emre Isik}
\email{isikf17@itu.edu.tr}
\author{Ali Demirci}
\email{demircial@itu.edu.tr}
\affiliation{Istanbul Technical University, Department of Mathematics Engineering,
Faculty of Science and Letters, Istanbul, 34469, Turkey}

\date{\today}

\begin{abstract}
Repeated decisions exhibit temporal structures such as persistence, direction-dependent switching, recurrent alternation, and abrupt transitions. We examine the generative sufficiency of a two-dimensional fast-slow dynamical system. The system combines a cubic fast equation with linear slow feedback and is analyzed through its equilibrium geometry, trace-determinant structure, equilibrium-fold loci, candidate Hopf boundaries, and singular critical manifold. An explicit observation operator projects continuous trajectories to a scalar signal and applies a binary readout, separating latent state-space dynamics from observable behavior. The analysis establishes a unique-equilibrium regime and, for suitable parameters, a three-equilibrium wedge with a central saddle. The outer equilibria are attracting only where their traces are negative. The analysis also identifies the simple-zero condition required for ordinary saddle-nodes, trace-zero positive-determinant spectral boundaries compatible with oscillatory instability, and the attracting and repelling branches of the critical manifold. Prescribed nonautonomous sweeps numerically illustrate direction-dependent switching (T1), transient episodic recurrent switching (T2), and an abrupt localized regime shift (T3). The attracting-equilibrium regime associated with prolonged state retention (T4) is characterized analytically. The resulting correspondence is intended as a test of generative sufficiency at the level of observable temporal organization, rather than as an identification or empirical validation of latent cognitive mechanisms.
\end{abstract}


\maketitle

\section{Introduction}
\label{sec:introduction}

Repeated decisions are temporally organized. Observable sequences retain
information about the timing and ordering of state changes, the recurrence of
switches, and the duration for which a selected state persists. This structure
is suppressed when trials are treated as independent observations. Empirical
and modeling work has instead shown that repeated-choice processes can evolve
over time, exhibit sequential dependencies, and contain low-dimensional or
regime-switching structure \cite{gunawan_time-evolving_2021,
kelly_response-time_2001,scherbaum_single_2022}. Persistence and choice inertia
have been observed across value-based, perceptual, and intertemporal settings
\cite{senftleben_stay_2021,senftleben_choice_2019,
akaishi_autonomous_2014,mcguire_decision_2012}. Other studies report
history-dependent responses and non-coincident transitions under comparable
external conditions \cite{koch_path_2009,thiel_hysteresis_2014,
hytonen_path_2014,jansen_hysteresis_2016}. Abrupt strategy changes and
oscillatory or fluctuating decision-related signals provide further examples
of behavior whose organization is temporal rather than trial-local
\cite{luwel_using_2001,gluth_deciding_2013,da_fonseca_mood_2023,
gheondeaeladi_bifurcation_2022}.

These phenomena are often studied through different tasks and explanatory models. Nevertheless, several are compatible with dynamical structures that can organize temporal behavior in qualitatively similar ways. Coexisting attractors can support path dependence; loss of an attracting state can be associated with a rapid transition; oscillatory dynamics can support recurrent threshold crossings; and a stable attractor can support prolonged residence. Attractor models and reduced dynamical descriptions are therefore natural tools for studying decision-related sequences \cite{albantakis_changes_2011,bitzer_bayesian_2015,schoemann_high-_2020}. The question addressed here is deliberately narrower than whether such models are psychologically complete: can one parsimonious dynamical structure generate multiple temporal signatures documented in repeated-decision behavior once its latent trajectories are mapped through an explicit observation operator?

The fast-slow distinction is used here as mathematical motivation rather than
as a direct assignment of cognitive function. Although the contrast between
fast and slow processes has a prominent place in descriptions of judgment
\cite{kahneman_thinking_2012}, the available experimental literature does not
provide temporally resolved latent-state observations that would allow the
state variables of the present dynamical system to be identified with specific
cognitive processes. Accordingly, neither state variable is assigned a
cognitive identity. The system is instead a modified
FitzHugh--Nagumo-type structure \cite{fn_model}: a cubic fast subsystem is coupled to a linear slow feedback variable. We denote the corresponding latent state variables by $s$ and $d$, where $s$ is the fast dynamical variable and $d$ is the slower feedback variable. These symbols are used purely as dynamical coordinates and are not assigned specific cognitive identities. This low-dimensional choice can support equilibrium multiplicity, bifurcation-driven transitions, local
oscillatory instability, and slow-fast motion without introducing a larger cognitive architecture.

A further issue concerns the distinction between latent dynamical states and
observable behavior. Phase portraits, equilibria, and bifurcation curves describe latent model states, whereas behavioral records generally contain discrete or otherwise restricted outputs. A property of state space is not automatically visible after measurement. For example, bistability does not produce an observed hysteresis loop unless the experimental protocol visits both branches and the observation rule registers the associated transitions.
We therefore introduce a separate observation operator that projects trajectories to a scalar signal and applies a binary readout. All signatures are defined on that output rather than by direct inspection of $(s,d)$.

The article makes three contributions. First, it provides a compact mathematical organization of the fast-slow system into equilibrium, stability, spectral, and critical-manifold results. Second, it makes the restriction from latent dynamics to observed behavior explicit through an observation operator.
Third, it defines four observable signatures: T1, direction-dependent switching; T2, recurrent switching; T3, an abrupt localized shift; and T4, prolonged state retention. Prescribed protocols numerically illustrate T1--T3, whereas T4 is associated analytically with convergence to an attracting equilibrium. The tags describe different properties of an output sequence and are not asserted to be mutually exclusive mechanism labels. In particular, T1 and T3 can both be compatible with equilibrium-fold-related dynamics, while T1 requires a comparison of forward
and reverse switching locations and T3 concerns the temporal localization of an individual latent transition relative to the imposed one-way protocol.

The central claim is consequently one of generative sufficiency. The model and observation operator are used to determine whether a common low-dimensional dynamical architecture can generate observable temporal signatures that have independently been documented in the decision-making literature. The correspondence is therefore sought at the level of temporal organization
rather than latent cognitive identity. The model and observation operator numerically illustrate T1--T3 and analytically characterize an attracting-equilibrium regime associated with T4. This does not imply that fast-slow dynamics are necessary, that the proposed system is unique, that its variables are identifiable cognitive processes, or that the numerical
illustrations constitute empirical validation of the model.

\section{Mathematical organization of the fast-slow system}
\label{sec:mathematical-organization}
Before introducing the observation layer, we first characterize the latent dynamical system independently of any behavioral readout. The purpose of this section is to identify the equilibrium, stability, bifurcation, and slow--fast
structures that can support the temporal signatures studied later. These results concern the state-space organization of the model itself and should therefore be distinguished from the observable signatures generated only after a protocol and observation operator are specified. The dynamical properties developed below are standard consequences of the FitzHugh--Nagumo structure and are included here primarily to establish the mathematical framework required for the subsequent observation-based analysis \cite{fn_model}. In particular, the equilibrium geometry, local stability, spectral boundaries, and slow--fast organization are specialized to the parameterization adopted in Eqs.~(1) and~(2).

\subsection{Model and notation}
\label{subsec:model}

We consider
\begin{align}
\varepsilon \dot{s}
  &=s-\frac{s^3}{3}-d+W,
  \label{eq:model-fast}\\
\dot{d}
  &=\rho(s+\mu-\nu d),
  \label{eq:model-slow}
\end{align}
where $0<\varepsilon\ll1$, $\rho,\nu>0$, and
$\mu,W\in\mathbb R$. Equations~(1) and~(2) constitute a modified FitzHugh--Nagumo (FHN) fast--slow system \cite{fn_model}, with a cubic fast subsystem coupled to a linear slow feedback equation. The variable $s$ is fast relative to $d$.
The parameter $W$ is an external control input, $\mu$ translates the slow nullcline, $\nu$ controls its slope and feedback strength, and $\rho$ sets the adjustment rate of $d$. These roles are dynamical rather than psychological.

The nullclines are
\begin{equation*}
d=s-\frac{s^3}{3}+W,
\qquad
d=\frac{s+\mu}{\nu}.
\end{equation*}
Their intersections satisfy
\begin{equation}
F(s;\mu,\nu,W)
=s-\frac{s^3}{3}-\frac{s+\mu}{\nu}+W=0.
\label{eq:F}
\end{equation}
For each real root $s_i^*$, the corresponding slow coordinate is
\begin{equation*}
d_i^*=\frac{s_i^*+\mu}{\nu}.
\end{equation*}
Whenever three distinct roots exist, they are labelled in increasing order,
\begin{equation*}
s_0^*<s_1^*<s_2^*,
\qquad
E_i=(s_i^*,d_i^*),
\quad i=0,1,2.
\end{equation*}

\subsection{Equilibrium geometry and multiplicity}
\label{subsec:equilibrium-geometry}
Having characterized the equilibrium geometry, we next examine the local dynamics in a neighborhood of these equilibria. Linearization provides the stability information needed to distinguish the dynamically admissible equilibrium branches and to locate parameter boundaries at which the local spectral structure changes. These results also provide the local dynamical foundation for the temporal regimes considered in the subsequent sections.
Multiple roots of Eq.~\eqref{eq:F} satisfy $F=F_s=0$. Since
\begin{equation*}
F_s(s)=1-s^2-\frac{1}{\nu},
\end{equation*}
real tangencies require $\nu>1$. Define
\begin{equation}
\alpha=\sqrt{1-\frac{1}{\nu}},
\qquad
W_\pm(\mu)=\frac{\mu}{\nu}
\mathbin{\pm}\frac{2}{3}
\left(1-\frac{1}{\nu}\right)^{3/2}.
\label{eq:fold-curves}
\end{equation}
In Eq.~\eqref{eq:fold-curves}, the compact notation means
$W_-(\mu)=\mu/\nu-\frac{2}{3}(1-1/\nu)^{3/2}$ and
$W_+(\mu)=\mu/\nu+\frac{2}{3}(1-1/\nu)^{3/2}$. The separation between the two equilibrium-fold curves is therefore
\begin{equation}
\Delta W_{\mathrm{fold}}
=
W_+(\mu)-W_-(\mu)
=
\frac{4}{3}
\left(1-\frac{1}{\nu}\right)^{3/2},
\qquad \nu>1.
\label{eq:fold-separation}
\end{equation}
This quantity characterizes the separation of the two fold loci in the frozen
equilibrium problem and should not, by itself, be interpreted as an observed
hysteresis width.

\begin{theorem}[Equilibrium multiplicity]
\label{thm:equilibrium-multiplicity}
If $\nu\leq1$, Eq.~\eqref{eq:F} has one real root for every
$(\mu,W)$. If $\nu>1$, it has three distinct real roots precisely when
\begin{equation*}
W_-(\mu)<W<W_+(\mu).
\end{equation*}
Two roots coalesce on either boundary, and the equilibrium is unique outside
the closed wedge.
\end{theorem}

\begin{figure}[h]
\centering
\includegraphics[
  width=0.78\linewidth,
  height=0.42\textheight,
  keepaspectratio
]{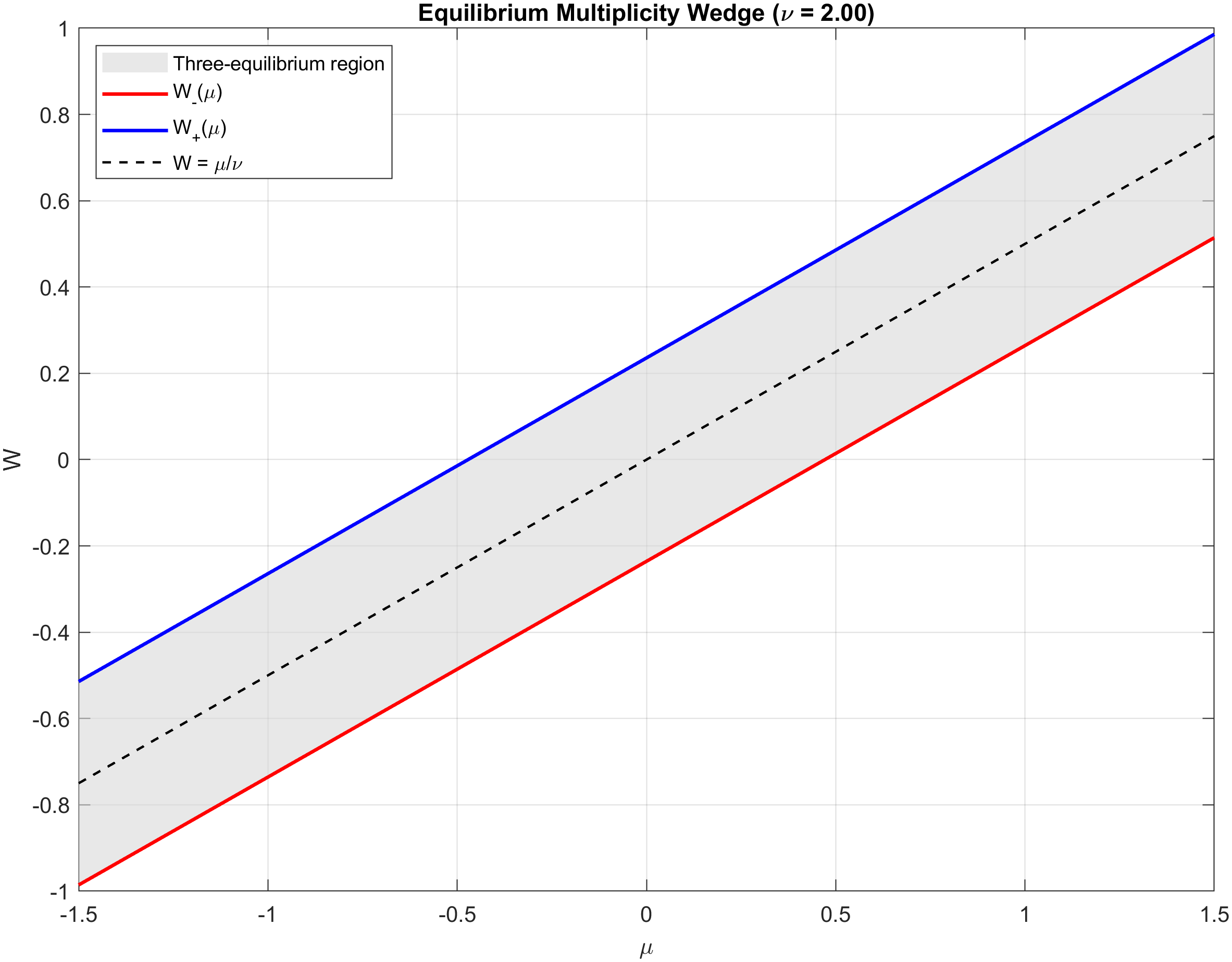}
\caption{Equilibrium multiplicity in the $(\mu,W)$ plane for $\nu=2$.
The curves $W_-(\mu)$ and $W_+(\mu)$ bound the shaded
three-equilibrium wedge, while the dashed center line is $W=\mu/\nu$.
Two scalar equilibria coalesce on each boundary; outside the wedge the
equilibrium is unique. The figure establishes multiplicity, not the stability
of the outer equilibria.}
\label{fig:wedge-region}
\end{figure}

This result follows equivalently from nullcline tangency or from the
discriminant of the depressed cubic associated with Eq.~\eqref{eq:F}. It
identifies multiplicity, but not by itself the stability of the coexisting
equilibria. When $\nu=1$ and $W=\mu$, the equilibrium is still unique, but
Eq.~\eqref{eq:F} reduces to $-s^3/3=0$ and therefore has a triple root at
$s=0$. This is a nonhyperbolic degenerate case rather than an ordinary
unique-hyperbolic-equilibrium case. The equilibrium-multiplicity partition
for the representative case $\nu=2$ is shown in Fig.~\ref{fig:wedge-region}.

\subsection{Stability and local bifurcations}
\label{subsec:local-stability}
The local analysis above characterizes the behavior of the system in the neighborhood of its equilibria. The small parameter $\varepsilon$, however, introduces an additional geometric structure associated with the separation
of fast and slow timescales. We therefore complement the equilibrium-based analysis by examining the singular limit and the resulting critical-manifold geometry, which organizes slow evolution and rapid latent transitions away from equilibrium.

At an equilibrium $E_i$, the Jacobian is
\begin{equation*}
J(E_i)=
\begin{pmatrix}
\dfrac{1-(s_i^*)^2}{\varepsilon} & -\dfrac{1}{\varepsilon}\\[2mm]
\rho & -\rho\nu
\end{pmatrix},
\end{equation*}
with
\begin{align}
\operatorname{tr}J(E_i)
  &=\frac{1-(s_i^*)^2}{\varepsilon}-\rho\nu,
  \label{eq:trace}\\
\det J(E_i)
  &=\frac{\rho}{\varepsilon}
    \left[1-\nu\left(1-(s_i^*)^2\right)\right].
  \label{eq:determinant}
\end{align}
These expressions provide the basis for distinguishing the stability of the
coexisting equilibria and for identifying the local spectral conditions
associated with the bifurcation boundaries considered below.

\begin{lemma}[Stability structure]
\label{lem:stability}
Within the three-equilibrium wedge, $E_1$ is a saddle. The outer
equilibria $E_0$ and $E_2$ have positive determinant; their local stability
is therefore determined by the signs of Eqs.~\eqref{eq:trace} and
\eqref{eq:determinant}. An outer
equilibrium is locally asymptotically stable only where its trace is negative.
Consequently, the wedge establishes equilibrium multiplicity and a central
saddle; it is a bistable region only where both outer equilibria satisfy the
negative-trace condition.
\end{lemma}

\begin{theorem}[Equilibrium-fold loci and saddle-node classification]
\label{thm:saddle-node}
For $\nu>1$, the curves $W_-(\mu)$ and $W_+(\mu)$ are equilibrium-fold
loci: two roots of Eq.~\eqref{eq:F} coalesce at $s=\pm\alpha$. The scalar
tangency satisfies $F_{ss}(\pm\alpha)=\mp2\alpha\neq0$ and $F_W=1$.
At the corresponding planar equilibrium, the zero eigenvalue is simple only
if
\begin{equation}
1-\varepsilon\rho\nu^2\neq0.
\label{eq:simple-zero}
\end{equation}
When Eq.~\eqref{eq:simple-zero} holds and the standard parameter
transversality and center-manifold nondegeneracy conditions are satisfied, the
planar equilibrium is an ordinary saddle-node. If
$1-\varepsilon\rho\nu^2=0$, the equilibrium has a double-zero degeneracy and
is not an ordinary saddle-node.
\end{theorem}

The trace vanishes at
\begin{equation*}
s_H^{(\pm)}
=\pm\sqrt{1-\varepsilon\rho\nu},
\end{equation*}
provided $\varepsilon\rho\nu<1$. Substitution into the equilibrium relation
gives
\begin{equation}
W_H^{(\pm)}(\mu)
=\frac{s_H^{(\pm)}+\mu}{\nu}
-s_H^{(\pm)}
+\frac{(s_H^{(\pm)})^3}{3}.
\label{eq:hopf-loci}
\end{equation}
These curves identify the parameter locations at which the equilibrium spectrum can become purely imaginary, providing the spectral basis for the candidate Hopf boundaries characterized below.
\begin{theorem}[Hopf spectral boundaries]
\label{thm:hopf}
Assume $\varepsilon\rho\nu<1$ and
$1-\varepsilon\rho\nu^2>0$. On the curves in
Eq.~\eqref{eq:hopf-loci}, the relevant equilibrium satisfies
$\operatorname{tr}J=0$ and $\det J>0$, and hence has a purely imaginary
eigenvalue pair. These curves are candidate Hopf boundaries, or Hopf spectral
boundaries. A Hopf bifurcation may occur there if the parameter crossing is
transverse and the first Lyapunov coefficient is nonzero.
\end{theorem}

The equilibrium-fold and Hopf spectral curves identify different local
changes. Crossing an equilibrium fold changes the number of equilibria.
A transverse crossing of a Hopf spectral boundary changes the sign of the trace without
changing that count. The spectral calculation does not establish a
nondegenerate Hopf bifurcation, the emergence or stability of a periodic
orbit, or recurrent threshold crossings. It identifies parameter regions
compatible with oscillatory instability. Proximity between the two types of
curve can place equilibrium multiplicity and oscillatory-instability changes
on nearby parameter scales without making them the same event.

\subsection{Slow-fast geometry}
\label{subsec:slow-fast}
The preceding analysis concerns the equilibrium and local spectral structure of the full system. We now turn to its singularly perturbed geometry, which provides a complementary description of the separation between slow evolution
and rapid latent transitions when $0<\varepsilon\ll1$. This viewpoint is particularly useful for identifying the attracting and repelling portions of the critical manifold and the fold points at which normal hyperbolicity is lost.
Setting $\varepsilon=0$ in the fast equation defines the critical manifold
\begin{equation*}
\mathcal C_0=
\left\{(s,d)\in\mathbb R^2:
d=s-\frac{s^3}{3}+W\right\}.
\end{equation*}
The fast eigenvalue along $\mathcal C_0$ is
$\lambda_f=1-s^2$.
The sign of $\lambda_f$ determines the normal stability of the corresponding branches of $\mathcal{C}_0$ and identifies the points at which normal hyperbolicity is lost. This structure is summarized in Lemma~\ref{lem:critical-manifold}.
\begin{lemma}[Critical-manifold structure]
\label{lem:critical-manifold}
The branches of $\mathcal C_0$ with $s<-1$ and $s>1$ are normally
attracting; the branch with $-1<s<1$ is normally repelling. Normal
hyperbolicity is lost at
\begin{equation}
F_-=\left(-1,W-\frac{2}{3}\right),
\qquad
F_+=\left(1,W+\frac{2}{3}\right).
\label{eq:critical-folds}
\end{equation}
\end{lemma}

\begin{figure}[p]
\centering
\includegraphics[
  width=\linewidth,
  height=0.80\textheight,
  keepaspectratio
]{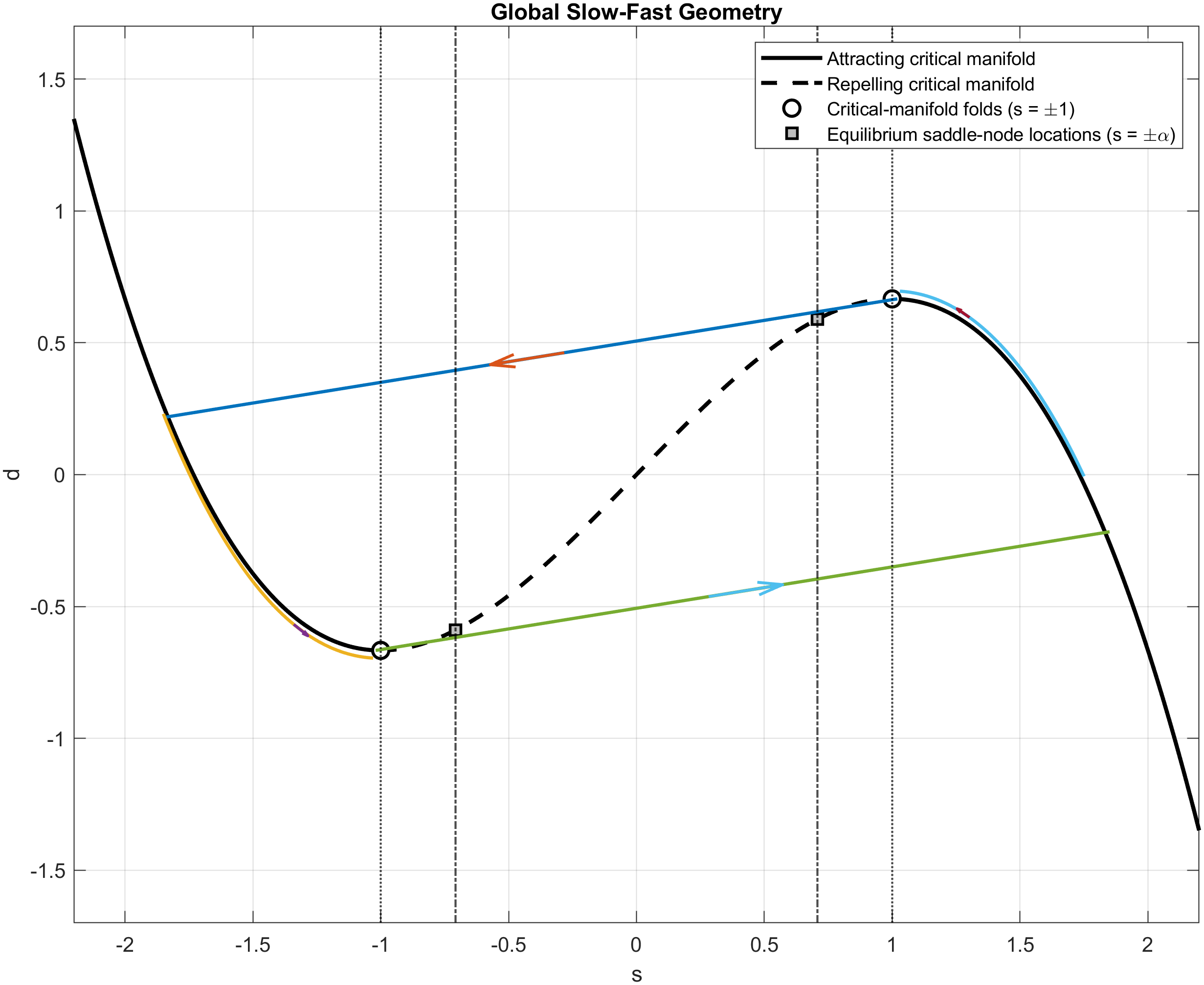}
\caption{Slow-fast geometry in the $(s,d)$ plane. Solid and dashed black
segments denote normally attracting and repelling portions of the critical
manifold. Open circles mark its folds at $s=\pm1$, and squares mark the
equilibrium-fold locations at $s=\pm\alpha$. The latter are ordinary planar
saddle-nodes only when the zero eigenvalue is simple and the standard
nondegeneracy conditions hold. The colored path illustrates geometry
compatible with slow drift and fast excursions; it does not prove a global
periodic orbit, relaxation cycle, or canard.}
\label{fig:global-slowfast-geometry}
\end{figure}

The folds in Eq.~\eqref{eq:critical-folds} are not the equilibrium-fold
locations at $s=\pm\alpha$. The former belong to the singular geometry of the
fast subsystem; the latter arise from tangency of the fast and slow nullclines
and are ordinary planar saddle-nodes only under the
conditions in Theorem~\ref{thm:saddle-node}. For a frozen, time-independent
value of $W$, evolution on the critical manifold away from $s=\pm1$ is
governed by
\begin{equation}
(1-s^2)\dot{s}
=\rho\left[
s+\mu-\nu\left(s-\frac{s^3}{3}+W\right)
\right].
\label{eq:reduced-flow}
\end{equation}
The reduced flow, together with the critical-manifold geometry, separates slow drift from rapid transitions. The geometry is compatible with hysteretic switching and,
under additional global matching conditions, with relaxation oscillations and canard-type behavior. The present article does not promote those conditional global mechanisms to proved existence claims. Under the nonautonomous sweep
protocols used below, $W=W(t)$ is prescribed; Eq.~\eqref{eq:reduced-flow} is
therefore a frozen-parameter relation and is not applied as a time-dependent
reduced equation. The distinction between critical-manifold folds and
equilibrium-fold locations is illustrated in Fig.~\ref{fig:global-slowfast-geometry}.

Table~\ref{tab:analytical-results} summarizes the analytical results and the
conditions required before they can be connected to observable consequences.

\section{From latent dynamics to observable signatures}
\label{sec:observation}
The analysis above characterizes the latent state-space organization of the model, but the behavioral signatures considered in this study are defined at the level of observable output. We therefore introduce an explicit observation
layer that separates the underlying continuous dynamics from the restricted signals available to an observer. This distinction provides the basis for the operational definitions of T1--T4 used below.

\subsection{Observation operator}
\label{subsec:observation-operator}

The observation layer is the composition
\begin{equation*}
\mathcal O=\mathcal R\circ\Pi,
\end{equation*}
where
\begin{equation}
z(t)=\Pi(s(t),d(t))=s(t)-\kappa d(t)-\theta
\label{eq:projection}
\end{equation}
and
\begin{equation}
y(t)=\mathcal R(z(t))=
\begin{cases}
+1, & z(t)>0 \\
-1, & z(t)\leq0.
\end{cases}
\label{eq:readout}
\end{equation}
The demonstrations use the baseline choice $\kappa=\theta=0$, so that $y(t)$ is determined by the sign of $s(t)$. Retaining the general projection in Eq.~\eqref{eq:projection} makes the measurement restriction explicit: different projections or thresholds could render different parts of the same
trajectory observable. For a fixed latent trajectory, changing $(\kappa,\theta)$ does not alter the underlying dynamics; it changes only which portions of that trajectory are
registered as threshold-crossing events. Consequently, the dynamical structures identified in Section~\ref{sec:mathematical-organization} are independent of the observation parameters, whereas the resulting binary signature need not be. The baseline choice used below should therefore be understood as one explicit readout of the model-generated trajectories rather
than as a claim that the corresponding signatures are invariant under all observation rules.

\begin{figure}[p]
\centering
\includegraphics[
  width=\linewidth,
  height=0.80\textheight,
  keepaspectratio
]{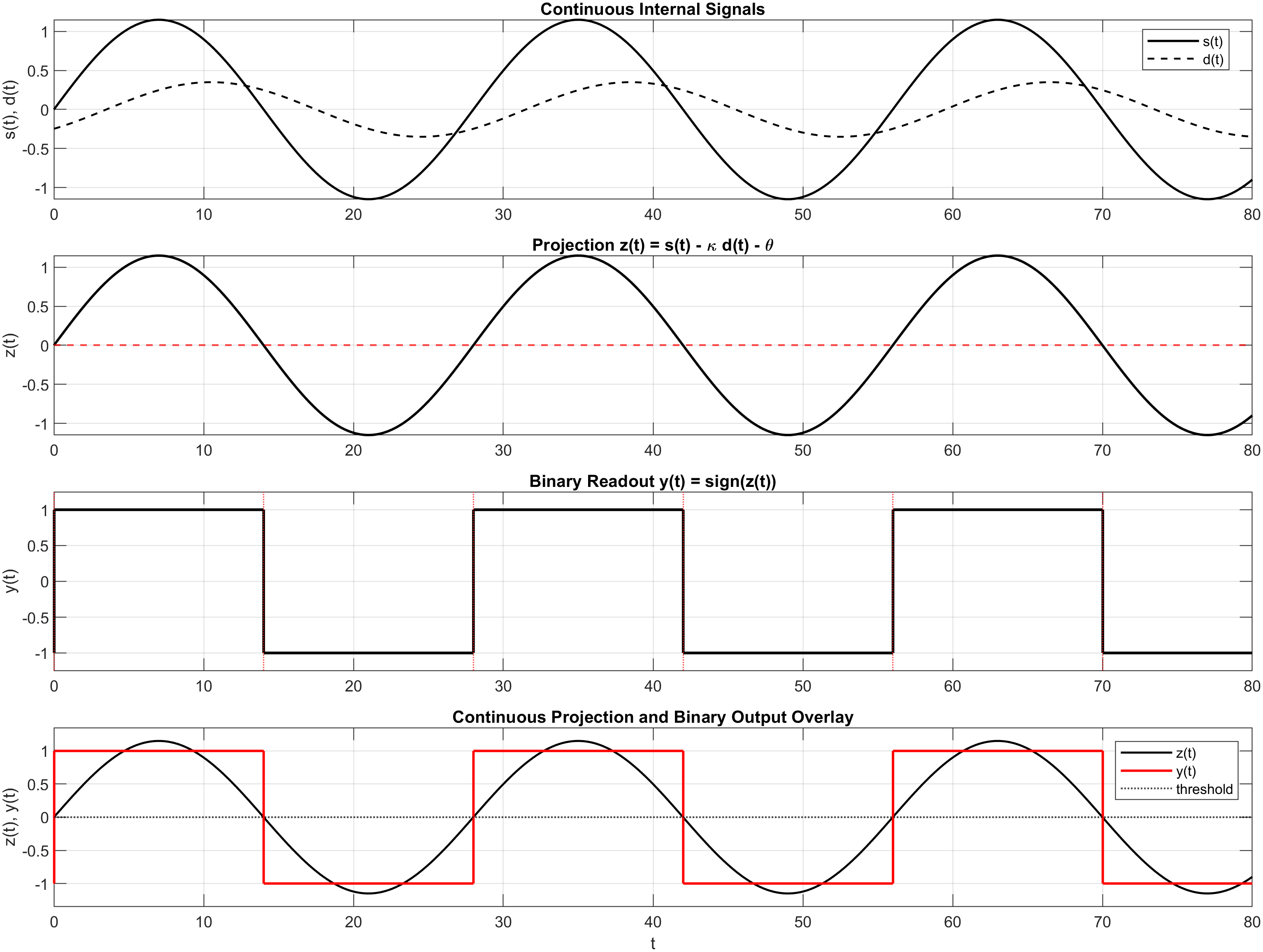}
\caption{Schematic illustration of the observation operator. Representative continuous signals $s(t)$ and $d(t)$ are projected to $z(t)=s(t)-\kappa d(t)-\theta$ and thresholded to obtain the binary readout $y(t)$. The lower overlay shows that changes in $y(t)$ occur at zero crossings of $z(t)$. The operator preserves transition ordering while
discarding most state-space information.}
\label{fig:observation-pipeline}
\end{figure}

The information reduction induced by this mapping is illustrated
schematically in Fig.~\ref{fig:observation-pipeline}.

\subsection{Observable descriptors and signature taxonomy}
\label{subsec:taxonomy}

Let $t_k$ denote consecutive switching times of $y(t)$. Dwell intervals are
\begin{equation*}
\tau_k=t_{k+1}-t_k.
\end{equation*}
Descriptive output properties also use transition ordering, the number and
recurrence of switches, direction-dependent switching locations, and
persistence over a finite observation window.

Table~\ref{tab:signature-taxonomy} formalizes the four tags. T1--T4 remain
distinct even where they are supported by related dynamics. T1 is defined by a paired, directional comparison; T2 by temporal recurrence; and T3 by one temporally localized latent transition relative to the imposed protocol. T4 is defined primarily by retention rather than transition.

\section{Model-generated signature demonstrations}
\label{sec:demonstrations}
The preceding section defined the observable signatures independently of any particular trajectory realization. We now examine how these operational forms can arise from the same latent dynamical system under prescribed protocols and
a fixed observation rule. The purpose of the demonstrations is not parameter estimation or empirical validation, but to establish explicit model-generated realizations of the signatures within the analytical structures identified
above.

\subsection{Demonstration protocol}
\label{subsec:protocol}

The present study evaluates the model in Eqs.~\eqref{eq:model-fast} and \eqref{eq:model-slow} through synthetic trajectories. T1 uses a prescribed forward-reverse nonautonomous
protocol $W=W(t)$, T2 uses a prescribed local parameter sweep $W=W(t)$ through a region compatible with oscillatory instability, and T3 uses a prescribed monotonic ramp $W=W(t)$. The trajectories are passed through Eqs.~\eqref{eq:projection}--\eqref{eq:readout}, and the reported evidence is read from the latent trajectory and binary output. The equilibrium, stability, and equilibrium-fold results in Section~\ref{sec:mathematical-organization} use frozen values of $W$. During a finite-rate sweep, an observed transition need not occur at a static equilibrium-fold location. The trajectory is generated by a nonautonomous protocol, and its subsequent crossing of the observation threshold is distinct
from the fold condition of the frozen equilibrium problem. Accordingly, the observed switching location should not be identified with a static fold location.

\subsection{T1: direction-dependent switching}
\label{subsec:T1}

T1 asks whether the observed transition location depends on the direction of
control-parameter variation. In the prescribed protocol $W=W(t)$, write
$W_{\mathrm{sw}}^\uparrow$ and $W_{\mathrm{sw}}^\downarrow$ for corresponding
observed switching locations during increasing and decreasing sweeps. The
observable criterion is
\begin{equation}
W_{\mathrm{sw}}^\uparrow\neq W_{\mathrm{sw}}^\downarrow.
\label{eq:T1}
\end{equation}
Empirical studies of hysteresis and path dependence, in which current responses depend on the prior trajectory through the stimulus or demand space, provide behavioral motivation for this observable criterion \cite{farrell_hysteresis_1999,thiel_hysteresis_2014,
jansen_hysteresis_2016,hytonen_path_2014}.

Under the prescribed forward--reverse protocol, T1 is identified at the observation level by the directional separation defined in
Eq.~\eqref{eq:T1}. The numerical trajectory provides an explicit realization of this criterion: the state follows different paths during the increasing and decreasing portions of the sweep, and the binary readout switches at distinct values of $W$. The relevant dynamical background is supplied by a region in
which attracting outer branches coexist with the equilibrium-fold geometry, but this structure is only a compatibility condition. Equilibrium multiplicity, or even the presence of equilibrium folds, does not by itself constitute observable hysteresis.

\begin{figure}[p]
\centering
\includegraphics[
  width=\linewidth,
  height=0.80\textheight,
  keepaspectratio
]{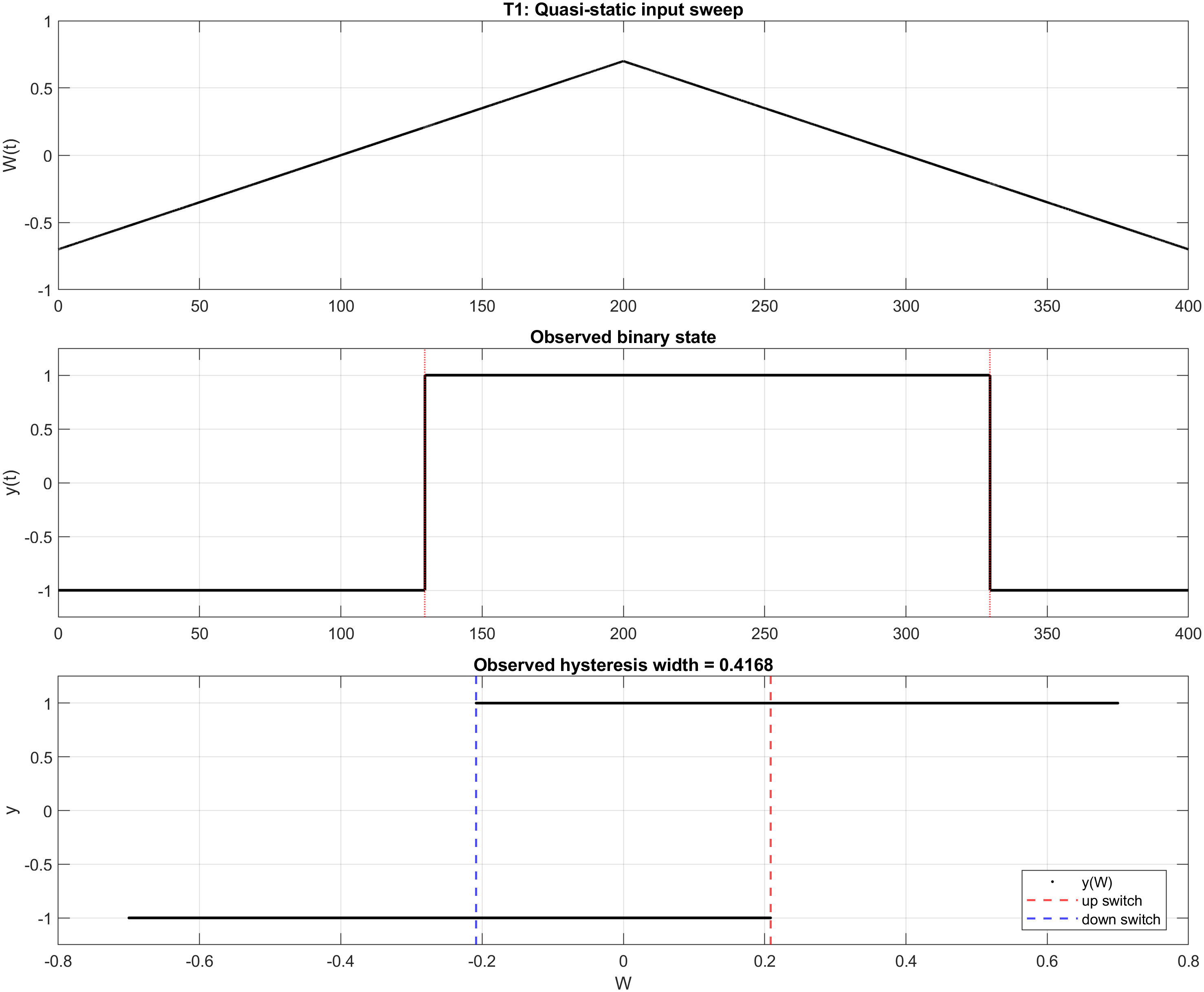}
\caption{T1: direction-dependent switching under a prescribed
forward-reverse nonautonomous protocol $W=W(t)$. The upper panel shows the sweep, the middle panel shows the binary readout, and the lower panel compares the observed switching locations. Their separation defines the observed hysteresis width for this
model-generated finite-rate trajectory; it is not identified with the separation between the static equilibrium folds.}
\label{fig:T1-hysteresis}
\end{figure}

This distinction is especially important because the two quantities involved belong to different levels of description. The observed switching separation is determined from threshold crossings of the protocol-driven trajectory, whereas the static fold separation $\Delta W_{\mathrm{fold}}$ in Eq.~\eqref{eq:fold-separation} is determined entirely by the frozen equilibrium geometry. The observed switching locations therefore need not
coincide with the static fold locations, since they additionally depend on the sweep-driven trajectory and the observation rule. T1 is consequently a protocol-conditioned observable signature rather than a direct label for equilibrium multiplicity or saddle-node geometry. Figure~\ref{fig:T1-hysteresis}
provides the corresponding numerical illustration.

\subsection{T2: recurrent switching}
\label{subsec:T2}

T2 is assigned when repeated sign changes of $y(t)$ form one or more
temporally organized episodes. The signature does not require uninterrupted
alternation throughout the observation window; transient episodic recurrence
followed or separated by constant-output intervals is admissible. Multiple
switches alone are insufficient without temporal localization or organization.
Fluctuating choice, hidden
changes of mind, and time-varying decision signals provide empirical
motivation for this signature \cite{gluth_deciding_2013,
da_fonseca_mood_2023,sullivan_indecision_2025,
scherbaum_single_2022}.

The candidate Hopf boundaries in Theorem~\ref{thm:hopf} identify regions compatible with oscillatory instability, while the slow--fast geometry can support recurrent motion when the required global conditions are satisfied. Neither analytical result, by itself, establishes a stable periodic orbit. For T2, the relevant criterion is instead observational: the prescribed local
nonautonomous sweep generates a trajectory whose projection crosses the observation threshold repeatedly in temporally organized episodes. The binary record therefore realizes recurrent switching, including intervals of
persistent output between successive switching episodes.

\begin{figure}[p]
\centering
\includegraphics[
  width=\linewidth,
  height=0.80\textheight,
  keepaspectratio
]{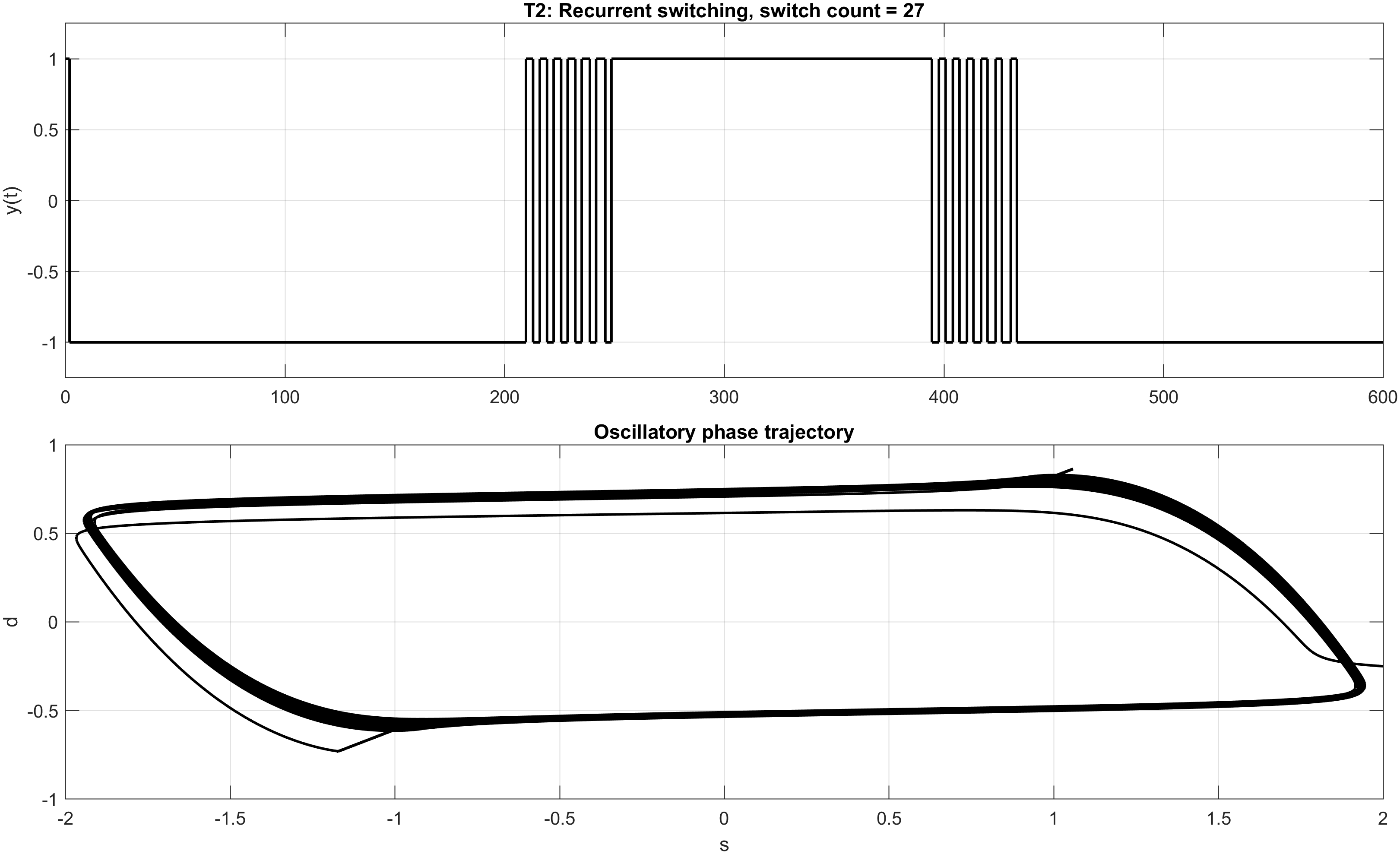}
\caption{T2: transient episodic recurrent switching under a prescribed local parameter sweep. The binary record exhibits localized episodes of recurrent switching separated by extended constant-output intervals. The lower panel shows the phase-space projection of the resulting nonautonomous trajectory
and should not be interpreted as evidence of uninterrupted sustained oscillation or of the existence or stability of a fixed-parameter periodic
orbit.}
\label{fig:T2-oscillatory}
\end{figure}

This distinction separates the observable signature from a particular oscillatory mechanism. The phase trace generated by the nonautonomous protocol is not a fixed-parameter limit cycle, and recurrent threshold crossings do not by themselves establish a Hopf bifurcation, a relaxation oscillation, or the
stability of a periodic orbit. T2 is consequently assigned from the temporal organization of the observed switching sequence rather than from a unique state-space mechanism. Figure~\ref{fig:T2-oscillatory} provides an explicit
numerical realization of this protocol-conditioned signature.

\subsection{T3: abrupt regime shift}
\label{subsec:T3}

T3 concerns a single observed binary switch associated with a temporally localized excursion of the latent trajectory relative to the imposed parameter ramp. The sign change in $y(t)$ registers the switch but does not by itself establish abruptness, because any ordinary threshold crossing changes
the binary output. The output is persistent on each side of the event, and no recurrent switching is required nearby. Abrupt strategy changes and change-point descriptions in decision tasks motivate this observable form
\cite{luwel_using_2001,van_rooij_what_2013}.

Under the prescribed monotonic protocol $W=W(t)$, the numerical trajectory provides an explicit realization of T3. The latent state undergoes a localized excursion over a time interval that is short relative to the gradual variation of the imposed control parameter, while the binary readout registers a single
switch and subsequently remains in the new observed state. The signature is therefore identified from the joint temporal organization of the latent transition and the observed output, rather than from the binary sign change alone. Within the present analysis, abruptness is used in this relative,
qualitative sense and is not introduced as a separately calibrated numerical index.

\begin{figure}[p]
\centering
\includegraphics[
  width=\linewidth,
  height=0.80\textheight,
  keepaspectratio
]{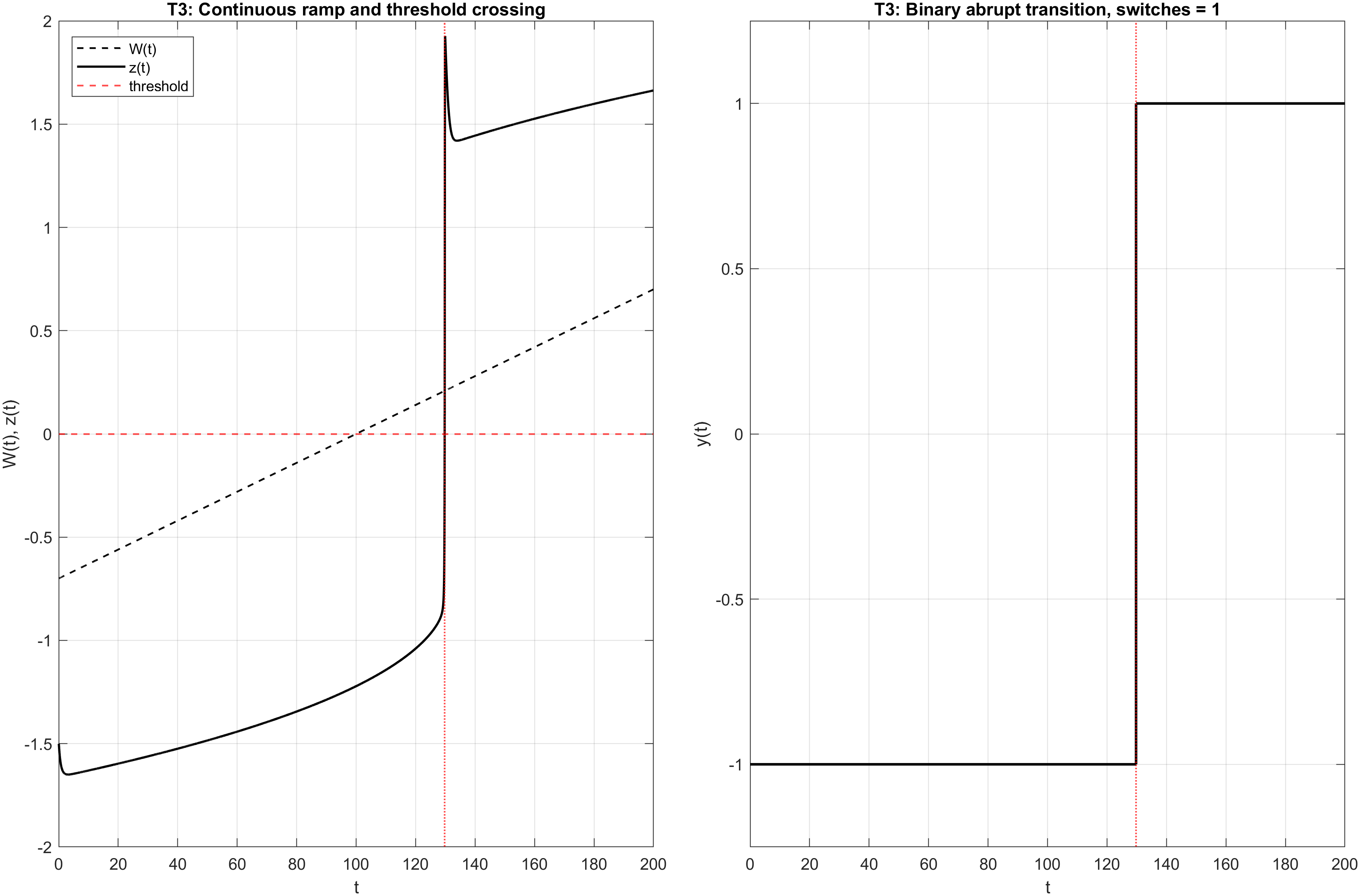}
\caption{T3: one observed binary switch under a prescribed monotonic protocol $W=W(t)$. The left panel shows a temporally localized excursion of the latent projected signal $z(t)$ relative to the gradual ramp, and the right panel shows the resulting switch in $y(t)$. Abruptness is a qualitative
trajectory-level diagnostic here, not a consequence of the binary sign change alone or a calibrated classification rule. The observed crossing is not identified automatically with a static equilibrium fold.}
\label{fig:T3-abrupt-shift}
\end{figure}

From a dynamical perspective, the T3 demonstration is organized by equilibrium-fold-related attractor loss under the prescribed monotonic ramp. As the control parameter is varied, the trajectory leaves the attracting regime and undergoes a rapid transition toward a different state-space region, producing the localized change observed in the output. Under a finite-rate protocol, however, the observed switching location need not coincide exactly with the corresponding static equilibrium-fold location, since it additionally depends on the driven trajectory and the observation threshold. T3 is therefore associated here with fold-related loss of an attracting state, while its observable definition remains the temporally localized transition and subsequent persistent output change. This also distinguishes T3 from T1: T1 compares direction-specific switching locations under a forward–reverse protocol, whereas T3 concerns a single localized transition under a one-way ramp. Figure~\ref{fig:T3-abrupt-shift} provides the corresponding numerical illustration.

\subsection{T4: prolonged state retention}
\label{subsec:T4}

T4 is characterized by prolonged state retention. Empirical findings on repetition, choice inertia, and persistent responding provide behavioral motivation for this observable form
\cite{alos-ferrer_inertia_2016,akaishi_autonomous_2014,
ashby_effect_2019,gershman_origin_2020,senftleben_stay_2021}.

At the observation level, T4 is assigned when $y(t)$ remains in the same state for an extended interval, with few or no switching events and dwell times that are long relative to the characteristic timescale of fast transitions. The defining property is therefore prolonged observable occupancy rather than a transition event.

An attracting equilibrium provides a sufficient dynamical regime for generating this signature. Let $E^*=(s^*,d^*)$ be an asymptotically stable equilibrium for fixed parameters, and suppose that the initial condition lies in its basin of attraction. Then the latent trajectory converges to $E^*$. Under the observation operator, define
\[
z^*=s^*-\kappa d^*-\theta .
\]
If $z^*\neq0$, the limiting projected state lies strictly on one side of the readout threshold. Hence there exists a sufficiently late time after which the projected trajectory remains on that same side of the threshold, and the binary output is constant.

The observation condition is essential. Asymptotic attraction in the latent state space does not by itself guarantee observable retention when the limiting projected state lies exactly on the threshold, $z^*=0$. By contrast, if $z^*\neq0$, convergence guarantees post-transient retention and therefore produces the long-dwell, low-switching structure that defines T4. The attracting-equilibrium argument thus provides an analytical sufficient condition for the observable dwell-time stability associated with this signature.

\subsection{Cross-signature organization}
\label{subsec:cross-signature}

The four signatures can now be compared within a common dynamical and observational framework. Their distinction does not rely on assigning a unique dynamical mechanism to each behavioral form. Instead, each signature is
specified by the combination of a compatible latent structure, an imposed protocol, and an observable temporal criterion. T1 is defined by a directional difference between switching locations under a forward--reverse sweep; T2 by temporally organized recurrence of threshold
crossings; T3 by a temporally localized latent transition relative to the imposed
one-way protocol, accompanied by a persistent observed state change; and T4 by
prolonged observable retention, characterized by long dwell intervals and few or
no switching events.

The level of evidence also differs across the four signatures. T1--T3 are established
here as explicit model-generated realizations under prescribed nonautonomous
protocols, whereas for T4 an attracting equilibrium together with the observation
condition provides an analytical sufficient condition for post-transient retention. 

The results of Section~\ref{sec:mathematical-organization} identify dynamical structures compatible with these signatures, but do not establish a one-to-one mapping between a bifurcation mechanism and an observable behavioral pattern. The comparative result is therefore that the same low-dimensional fast--slow system can generate distinct forms of observable temporal organization when the protocol and the information retained by the observation operator are taken into account.

\section{Discussion}
\label{sec:discussion}
The analytical results and model-generated demonstrations can now be considered together to clarify what the framework does, and does not, imply about observable decision-related dynamics. The discussion below focuses on the relation between latent dynamical structure and observable signatures, the information restriction imposed by the observation operator, and the interpretive scope of the resulting correspondence with the behavioral literature.
\subsection{One system, multiple observable signatures}
\label{subsec:one-system}

The principal result is structural: several organizations of observable
decision-related time series can be generated within one low-dimensional
model. The three-equilibrium wedge establishes multiplicity and a central
saddle; branch coexistence becomes bistability only where both outer
equilibria are attracting. Equilibrium-fold boundaries mark coalescence of
equilibria and become ordinary saddle-nodes only under the simple-zero and
standard nondegeneracy conditions. Hopf spectral boundaries identify regions
compatible with oscillatory instability but do not prove periodic-orbit
emergence or stability. The critical manifold separates normally attracting
and repelling branches, while fast jumps and recurrent global motion remain
conditional on the full trajectory geometry. None of these structures
constitutes a behavioral signature by itself. A signature appears only after a
protocol selects a trajectory and the observation operator registers its
threshold crossings.

The cross-signature comparison also shows why the observable taxonomy should not be interpreted as a one-to-one classification of latent mechanisms. Different signatures can be compatible with overlapping dynamical structures, while the same latent organization can produce different observable forms under different protocols. The distinction among T1--T4 therefore lies in the temporal organization retained at the observation level, rather than in an assumption that each signature originates from a unique bifurcation or state-space mechanism.

\subsection{The observation operator as an information restriction}
\label{subsec:information-restriction}

The observation operator prevents a direct identification of state-space geometry with behavior. Equation~\eqref{eq:projection} maps two continuous
coordinates to one scalar and Eq.~\eqref{eq:readout} reduces that scalar to one bit at each time. Amplitude, distance to an equilibrium, most phase information, and much of the slow-fast geometry are discarded. Only the state of the threshold and the ordering of threshold crossings are retained.

This information loss has two implications. First, dynamical distinctions can be observationally equivalent. An oscillatory trajectory that remains on one side of $z=0$ produces no switching, whereas recurrent motion that crosses the
threshold can produce T2. Second, output distinctions can depend on the protocol. A bistable regime can yield T1 only when both sweep directions are observed and the trajectory registers separated switches; a one-way traversal of compatible fold geometry can instead produce a T3-like fast excursion. Observable behavioral organization thus belongs to the combined object \emph{dynamics + protocol + observation operator}, rather than to the latent phase portrait alone.

\subsection{Interpretive scope}
\label{subsec:interpretive-scope}

The framework complements, rather than replaces, cognitive and statistical
accounts. Existing research explains choice inertia through mechanisms such
as residual activity, learning, reward-complexity trade-offs, or switching
costs \cite{senftleben_choice_2019,akaishi_autonomous_2014,
gershman_origin_2020,ashby_effect_2019}. The present result lies at a different level: it shows that temporal patterns
resembling inertia, direction dependence, recurrence, and abrupt change can be generated within the proposed nonlinear dynamical framework under appropriate protocol and observation conditions. Such correspondence is structural and does not imply identification of the underlying psychological mechanism;
observed membership in T1--T4 therefore cannot by itself determine the latent mechanism responsible for a behavioral pattern. Empirical discrimination would require parameter
estimation, competing observation models, and predictions beyond the qualitative signature.

\subsection{Limitations}
\label{subsec:limitations}

The study has four main limitations.
\begin{enumerate}
\item The system is deterministic, whereas empirical decision sequences contain measurement noise and intrinsic variability.
\item No empirical dataset is fitted, so correspondence with the
decision-making literature is at the level of temporal form and dynamical compatibility.
\item The model-generated demonstrations of T1--T3 establish explicit existence examples under the prescribed protocols, but they do not establish the prevalence or robustness of these signatures across parameter space.
\item Relaxation oscillations, canards, and related global behaviors are invoked only as mechanisms compatible with the local slow-fast geometry under additional global assumptions; their global existence is not proved here.
\end{enumerate}

\section{Conclusion}
\label{sec:conclusion}

A two-dimensional fast-slow system, combined with an explicit observation
operator, provides a parsimonious framework for four forms of observable
temporal organization: direction-dependent switching, recurrent switching, an
abrupt localized shift, and prolonged state retention. The equilibrium-fold
analysis specifies when ordinary saddle-node classification is available, and
the trace-zero positive-determinant curves are retained as candidate Hopf
boundaries unless nonlinear bifurcation conditions are established. The
slow-fast and stability results identify structures compatible with the
observed signatures, while the observation layer prevents latent dynamics
from being treated as directly observed cognitive mechanisms.

The framework numerically illustrates T1--T3 and analytically characterizes the attracting-equilibrium regime associated with T4. It is a generative-sufficiency framework and does not constitute either an empirical validation of the model or an identification of the underlying psychological mechanism. A natural next step is therefore empirical. If temporally resolved experimental data capable of representing the evolution of decision-related behavior become available, the dynamical framework developed here could be confronted with observed time series rather than assessed solely through qualitative temporal signatures. Such data would permit the adequacy and limitations of the proposed fast--slow representation and its observation operator to be evaluated more directly, while potentially helping to formulate new experimentally testable questions concerning the temporal organization of decision-related behavior.

\section*{Ethics statement}

No human participants, animals, or empirical personal data were involved in this mathematical and simulation-based study.

\section*{Data availability}

No empirical dataset was collected or analyzed in this study. The results are based on analytical derivations and synthetic model trajectories generated within the mathematical framework described in the text.

\bibliography{tez}
\newpage
\section*{Tables}

\begin{table}[ht]
\caption{Analytical structure of the fast--slow model and its relation to
observable behavior.}
\label{tab:analytical-results}
\centering
\small
\renewcommand{\arraystretch}{1.18}
\begin{tabular}{|p{0.20\linewidth}|p{0.38\linewidth}|p{0.34\linewidth}|}
\hline
\textbf{Structure} &
\textbf{Dynamical result and conditions} &
\textbf{Observable implication} \\
\hline

Equilibrium multiplicity &
For $\nu>1$, three equilibria exist inside
$W_-<W<W_+$; outside this wedge the equilibrium is unique. The middle
equilibrium is a saddle. &
Coexisting observable states are possible only where both outer equilibria
are attracting. \\
\hline

Stability structure &
Inside the three-equilibrium wedge, $E_1$ is a saddle and
$E_0,E_2$ have positive determinant; their stability is determined by the
trace. &
Bistability requires both outer equilibria to have negative trace; multiplicity
alone is insufficient. \\
\hline

Equilibrium folds &
For $\nu>1$, two equilibria coalesce at $W=W_\pm$. Ordinary saddle-node classification additionally requires Eq.~\eqref{eq:simple-zero} and the standard nondegeneracy conditions. & Compatible with attractor loss, directional switching, or a localized latent transition when the relevant stability and protocol conditions are satisfied. \\
\hline

Hopf spectral boundaries &
For $\varepsilon\rho\nu<1$ and
$1-\varepsilon\rho\nu^2>0$, an outer equilibrium can have zero trace and
positive determinant. A Hopf bifurcation requires the additional nonlinear
conditions. &
Compatible with oscillatory instability; recurrent observable switching
additionally requires threshold-crossing recurrent motion. \\
\hline

Critical manifold &
For $0<\varepsilon\ll1$, the critical manifold has two normally attracting
outer branches, a repelling middle branch, and folds at $s=\pm1$. &
Compatible with slow residence and fast excursions; global jumps, relaxation
cycles, and canards require additional conditions. \\
\hline
\end{tabular}
\end{table}

\begin{table}[t]
\caption{Operational taxonomy of the four observable temporal signatures.}
\label{tab:signature-taxonomy}
\centering
\small
\renewcommand{\arraystretch}{1.20}
\begin{tabular}{|p{0.08\linewidth}|p{0.43\linewidth}|p{0.41\linewidth}|}
\hline
\textbf{Tag} &
\textbf{Operational definition} &
\textbf{Compatible dynamical structure} \\
\hline

T1 &
Direction-dependent switching under a prescribed forward--reverse sweep:
$W_{\mathrm{sw}}^\uparrow\neq W_{\mathrm{sw}}^\downarrow$. &
Attracting outer branches and equilibrium-fold geometry; the signature
requires comparison of the two sweep directions. \\
\hline

T2 &
Repeated threshold crossings organized into recurrent episodes under a
prescribed local parameter sweep. 
&
Local scillatory instability or global slow--fast recurrent motion. \\
\hline

T3 &
A persistent observed state change associated with a localized latent
excursion relative to the imposed ramp timescale. &
Compatible with conditional attractor loss or fold-related dynamics;
abruptness concerns the temporal localization of the latent transition
relative to the imposed ramp rather than the binary sign change alone. \\
\hline

T4 & Prolonged retention of one observable state, characterized by long dwell
intervals and few or no switching events.
&
An asymptotically stable equilibrium whose projected limiting state remains away
from the observation threshold provides a sufficient dynamical regime. \\
\hline
\end{tabular}
\end{table}
\clearpage
\section*{Figures}

For review purposes, all figures are collected after the references. They are
part of the main manuscript and are numbered consecutively in the order in
which they are cited in the text.

\end{document}